\documentclass[runningheads]{llncs}
\usepackage[T1]{fontenc}

\usepackage[table,xcdraw]{xcolor}
\usepackage{caption}
\usepackage{graphicx}
\usepackage{amsmath}

\usepackage{amsbsy}
\usepackage{amsfonts}
\usepackage{amssymb}
\usepackage{array}
\usepackage[misc]{ifsym}
\usepackage{hyperref}
\usepackage{subcaption}

\begin{document}

\title{Clustering disease trajectories in contrastive feature space for biomarker discovery in age-related macular degeneration}
\titlerunning{Clustering disease trajectories for temporal biomarker discovery in AMD}

\author{Robbie Holland \inst{1}$^\text{(\Letter)}$, Oliver Leingang \inst{2}, Christopher Holmes \inst{3}, Philipp Anders \inst{4}, Rebecca Kaye \inst{7}, Sophie Riedl \inst{2}, Johannes C. Paetzold \inst{1}, Ivan Ezhov \inst{8}, Hrvoje Bogunović \inst{2}, Ursula Schmidt-Erfurth \inst{2}, Lars Fritsche \inst{6}, Hendrik P. N. Scholl \inst{4, 5}, Sobha Sivaprasad \inst{3}, Andrew J. Lotery \inst{7}, Daniel Rueckert \inst{1,8}, Martin J. Menten \inst{1,8}}

\institute{BioMedIA, Imperial College London, London, United Kingdom \\\email{robert.holland15@ic.ac.uk} \and 
Laboratory for Ophthalmic Image Analysis, Medical University of Vienna \and 
Moorfields National Institute for Health and Care Biomedical Research Centre, Moorfields Eye Hospital, London, United Kingdom \and
Institute of Molecular and Clinical Ophthalmology Basel \and
Department of Ophthalmology, Universitat Basel, Basel, Switzerland \and
Department of Biostatistics, University of Michigan, Ann Arbor, United States \and
Clinical and Experimental Sciences, Faculty of Medicine, University of Southampton, Southampton, United Kingdom \and
Institute for AI and Informatics in Medicine, Technical University of Munich, Munich, Germany
}
\authorrunning{Holland \textit{et al}.}

\maketitle              
\begin{abstract}
Age-related macular degeneration (AMD) is the leading cause of blindness in the elderly. Current grading systems based on imaging biomarkers only coarsely group disease stages into broad categories and are unable to predict future disease progression. It is widely believed that this is due to their focus on a single point in time, disregarding the dynamic nature of the disease. In this work, we present the first method to automatically discover biomarkers that capture temporal dynamics of disease progression. Our method represents patient time series as trajectories in a latent feature space built with contrastive learning. Then, individual trajectories are partitioned into atomic sub-sequences that encode transitions between disease states. These are clustered using a newly introduced distance metric. In quantitative experiments we found our method yields temporal biomarkers that are predictive of conversion to late AMD. Furthermore, these clusters were highly interpretable to ophthalmologists who confirmed that many of the clusters represent dynamics that have previously been linked to the progression of AMD, even though they are currently not included in any clinical grading system.

\end{abstract}

\begin{figure}[h!]
\centering
\includegraphics[width=0.98\textwidth]{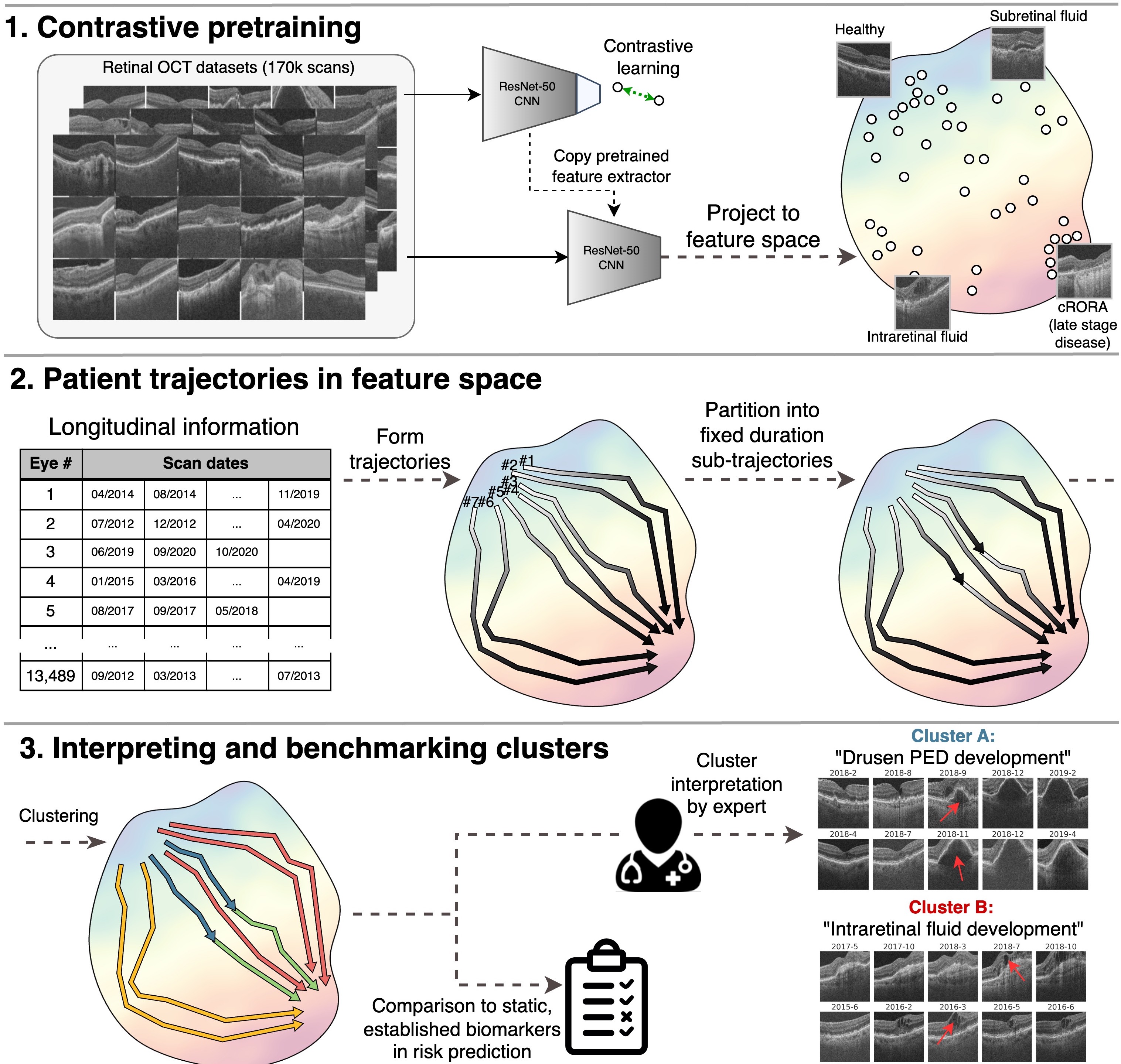}
\caption{Our method finds common patterns of disease progression in datasets of longitudinal images. We partition time series into sub-trajectories before introducing a clinically motivated distance function to clusters them in feature space. Time-dependent clusters are then assessed by expert ophthalmologists on their interpretability and ability to capture temporal dynamics related to the progression of AMD.} 
\label{fig:graphical_abstract}
\end{figure}

\section{Introduction}
Age-related macular degeneration (AMD) is the leading cause of blindness in the elderly, affecting nearly 200 million people worldwide \cite{wong2014global_s}. Patients with early stages of the disease exhibit few symptoms until suddenly converting to the late stage, at which point their central vision rapidly deteriorates \cite{mitchell2018age}. Clinicians currently diagnose AMD, and stratify patients, using biomarkers derived from optical coherence tomography (OCT), which provides high-resolution images of the retina. However, the widely adopted AMD grading system \cite{ferris2013clinical,sadda2018consensus}, which coarsely groups patients into broad categories for early and intermediate AMD, only has limited prognostic value for late AMD. Clinicians suspect that this is due to the grading system's reliance on static biomarkers that are unable to capture temporal dynamics which contain critical information for assessing progression risk.
\\
In their search for new biomarkers, clinicians have annotated known biomarkers in longitudinal datasets that monitor patients over time and mapped them against disease progression \cite{schlanitz2017drusen,steinberg2013longitudinal,chen2019longitudinal}. This approach is resource-intensive and requires biomarkers to be known \textit{a priori}. Others have proposed deep-learning-based methods to discover new biomarkers at scale by clustering OCT images or detecting anomalous features \cite{seebock2018unsupervised,waldstein2020unbiased,schlegl2019f}. However, these approaches neglect temporal relationships between images and the obtained biomarkers are by definition static and cannot capture the dynamic nature of the disease.\\

\noindent \textbf{Our contribution}:
In this work, we present a method to automatically discover biomarkers that capture temporal dynamics of disease progression in longitudinal datasets (see Figure \ref{fig:graphical_abstract}). At the core of our method is the novel strategy to represent patient time series as trajectories in a latent feature space. Individual progression trajectories are partitioned into atomic sub-sequences that encode transitions between disease states. Finally, we cluster these sub-trajectories using a newly introduced distance metric that encodes three distinct temporal criteria. The resulting clusters are treated and evaluated as proposals for temporal biomarkers, cataloguing population-level patterns of AMD progression.\\
Experimentally, we test our method on two large longitudinal retinal OCT datasets totalling 160,558 images from 7,912 patients. Expert ophthalmologists provide written interpretations by examining sets of longitudinal images from each cluster, describing existing or potentially new time-dependent biomarkers. Finally, we benchmark our clusters against the widely adopted imaging biomarker grading system, and find they often provide more accurate predictions of functional vision acuity and time to conversion to late AMD.

\section{Related work}

\noindent \textbf{Current AMD grading systems}: Ophthalmologists' current understanding of progression from early to late AMD largely involves drusen, which are subretinal lipid deposits. Drusen volume increases until suddenly stagnating and regressing, which often precedes the onset of late AMD \cite{schlanitz2017drusen}. The established AMD grading system stratifies early and intermediate stages solely by the size of drusen in a single OCT image \cite{bird1995international,klein2014harmonizing,ferris2005simplified,ferris2013clinical}. Late AMD is classified into either choroidal neovascularisation (CNV), identified by subretinal fluid, or geographic atrophy, signalled by progressive loss of photoreceptors and retinal thinning. The degree of atrophy can be staged using cRORA (complete retinal pigment epithelium and outer retinal atrophy), which measures the width in $\mu m$ of focal atrophy in OCT \cite{sadda2018consensus}. Grading systems derived from these biomarkers offer limited diagnostic value and little to no prognostic capability.\\

\noindent \textbf{Tracking evolution of known biomarkers}: Few research efforts have aimed at quantifying and tracking known AMD biomarkers, mostly drusen, over time \cite{steinberg2013longitudinal,schlanitz2017drusen}. More work has explored the disease progression of Alzheimer's disease (AD), which offers a greater array of quantitative imaging biomarkers, such as levels of tau protein and hippocampal volume. Young \textit{et al}. \cite{young2014data} fit an event-based model that rediscovers the order in which these biomarkers become anomalous as AD progresses. Vogel \textit{et al}. \cite{vogel2021four} find four distinct spatiotemporal trajectories for tau pathology in the brain. However, this only works if biomarkers are known \textit{a priori} and requires manual annotation of entire time series.\\

\noindent \textbf{Automated discovery of unknown biomarkers}:
Prior work for automated biomarker discovery in AMD explores the latent feature space of encoders trained for image reconstruction \cite{waldstein2020unbiased,seebock2018unsupervised}, segmentation \cite{zheng2020pathological} or generative adversarial networks \cite{schlegl2019f}. However, these neural networks are prone to overfit to their specific task and lose semantic information regarding the disease. Contrastive methods \cite{chen2020simple,grill2020bootstrap,zhao2022prognostic} encode invariance to a set of image transformations, which are uncorrelated with disease features, resulting in a more expressive feature space.\\
However, all aforementioned methods group single images acquired at one point in time, and in doing so neglect temporal dynamics. The one work that tackles this challenge, and the most related to ours, categorises the time-dependent response of cancer cells to different drugs, measured by the changing distance in contrastive feature space from healthy controls \cite{dmitrenko2021self}.

\section{Materials and methods}
\subsection{OCT image datasets}
\label{method:datasets}
We develop our method on an in-house retinal OCT dataset called the \textit{Development dataset}, collected from the Southampton Eye Unit, and test it on a second independent \textit{Unseen dataset} from Moorfields Eye Hospital. In both, images were acquired using Topcon 3D OCT devices (Topcon Corporation, Tokyo, Japan). After strict quality control, the \textit{Development dataset} consists of 46,496 scans of 6,236 eyes from 3,456 patients. Eyes were scanned 7.7 times over 1.9 years on average at irregular time intervals. The \textit{Unseen dataset} is larger, containing 114,062 scans of 7,253 eyes from 3,819 patients. Eyes were scanned 16.6 times over 3.5 years on average.\\
A subset of 1,031 longitudes was labelled using the established AMD grading protocols derived from known imaging biomarkers. Early AMD was characterised by small drusen between 63-125$\mu$m in diameter. We also recorded CNV, cRORA ($\geq 250 \mu$m and <1000$\mu$m), cRORA ($\geq1000 \mu$m) \cite{sadda2018consensus} and a healthy classification for cases with no visible biomarkers. Additionally, visual acuity scores, which measured the patient's functional quality of vision using a LogMAR chart, are available at 83,964 time points.

\subsection{Self-supervised feature space using contrastive learning}
\label{methods:cl}
We adapt BYOL \cite{grill2020bootstrap_s} for contrastive training of a ResNet50 (4x) model. As several of the contrastive transformations designed for natural images are inapplicable to medical images, such as solarisation, colour shift and greyscale, we use the set tailored for retinal OCT images by Holland \textit{et al}. \cite{DBLP:journals/corr/abs-2208-02529}. Models were trained on the entire dataset for 120,000 steps using the Adam optimiser with a momentum of 0.9 and a learning rate of $5 \cdot 10^{-4}$. After training, we first remove the final linear layer before projecting all labelled images to the feature space of 2048 dimensions.

\subsection{Extracting sub-trajectories via partitioning}
\label{methods:extracting}
Naively clustering whole time series of patients ignores two characteristics of longitudinal data. Firstly, individual time series are not directly comparable as patients enter and leave the study at different stages of their overall progression. Secondly, longer time series can record multiple successive transitions in disease stage. Inspired by TRACLUS \cite{lee2007trajectory}, the state of the art in trajectory clustering, we adapt their \textit{partition-and-group} framework by assuming that trajectories can be partitioned into a common set of \textit{sub-trajectories} that capture singular transitions between progressive states of the disease.\\
For each eye, we first form piecewise-linear trajectories by linking points in feature space that were derived from consecutively acquired OCT images. We then extract sub-trajectories by finding all sequences of images spanning $1.0\pm0.5$ years of elapsed time within each trajectory. Next, to avoid oversampling trajectories with a shorter time interval between images, we randomly sample at most one sub-trajectory in every 0.5-year time interval.

\begin{figure}[ht!]
\centering
\includegraphics[width=0.98\textwidth]{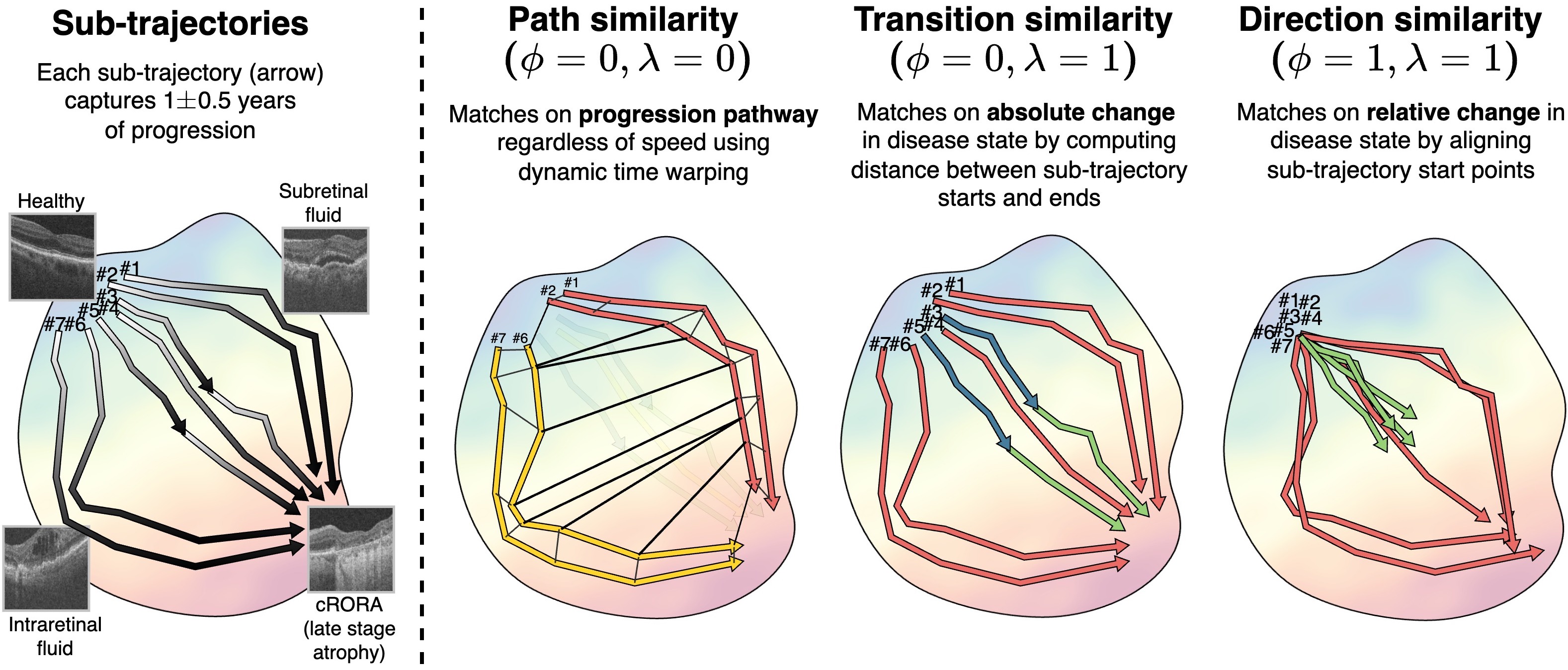}
\caption{Illustration of sub-trajectory distance functions which each encode temporal criteria for similarity (see Equation \ref{eq:dist}). We illustrate clusters assignments, denoted by colour, resulting from three combinations of $\phi$ and $\lambda$.}
\label{fig:method}
\end{figure}

\subsection{Sub-trajectory distance functions and clustering}
In order to find common patterns of disease progression among sub-trajectories we cluster them. To this end we introduce a new distance function between sub-trajectories that incorporates three distinct temporal criteria (see Figure \ref{fig:method}). The first, formulated in Equation \ref{eq:transition}, matches two sub-trajectories, $U$ and $V$, of patients who progress between the same start and end states:
\begin{align}
    D_{transition}(U,V) = \left\lVert U_{start} - V_{start}\right\rVert_2 + \left\lVert U_{end} - V_{end}\right\rVert_2\,.
    \label{eq:transition}
\end{align}
Since all sub-trajectories cover a similar temporal duration, $D_{transition}$ also differentiates between fast and slow progressors and stable periods of no progression. However, by ignoring intermediary images, this metric does not respect the disease pathway along which patients progress. To incorporate this, we include a second metric that measures path dissimilarity, calculated using dynamic time warping (DTW) \cite{sakoe1971dynamic,sakoe1978dynamic,cuturi2017soft}. DTW finds the optimal temporal alignment between two time series before computing their distance. This re-alignment allows us to match sub-trajectories that traverse the same disease states in the same order, irrespective of the rate of change between states. We combine $D_{transition}$ with DTW using a $\lambda$ coefficient so that the overall distance between $U$ and $V$ is
\begin{align}
  D_{path}(U,V) &= \lambda  \text{ } D_{transition}(U, V) + (1 - \lambda) \text{ DTW}(U, V)\,.
\label{eq:dist}
\end{align}
The third and final temporal criteria is to match time series that progress in the same relative direction, regardless of absolute disease states. We weight the contribution of this with $\phi$ in Equation \ref{eq:vector}:
\begin{equation}
      D_{subtraj}(U,V) = \phi  \text{ } D_{path} (U - U_{start}, V - V_{start})
  + (1 - \phi) \text{ } D_{path}(U, V)\,.
  \label{eq:vector}
\end{equation}

\subsubsection{Spectral clustering} 
As the non-linearity of $D_{subtraj}$ prohibits the use of k-means for clustering, we instead use spectral clustering \cite{von2007tutorial} to group similar sub-trajectories. Hereby, we construct an affinity matrix $\mathcal{A}$ encoding the negative of the distance $D_{subtraj}$ between all pairs of sub-trajectories. Using $\mathcal{A}$, we group sub-trajectories into $K$ clusters.

\begin{figure}[htp]
    \centering
    \includegraphics[width=0.99\linewidth]{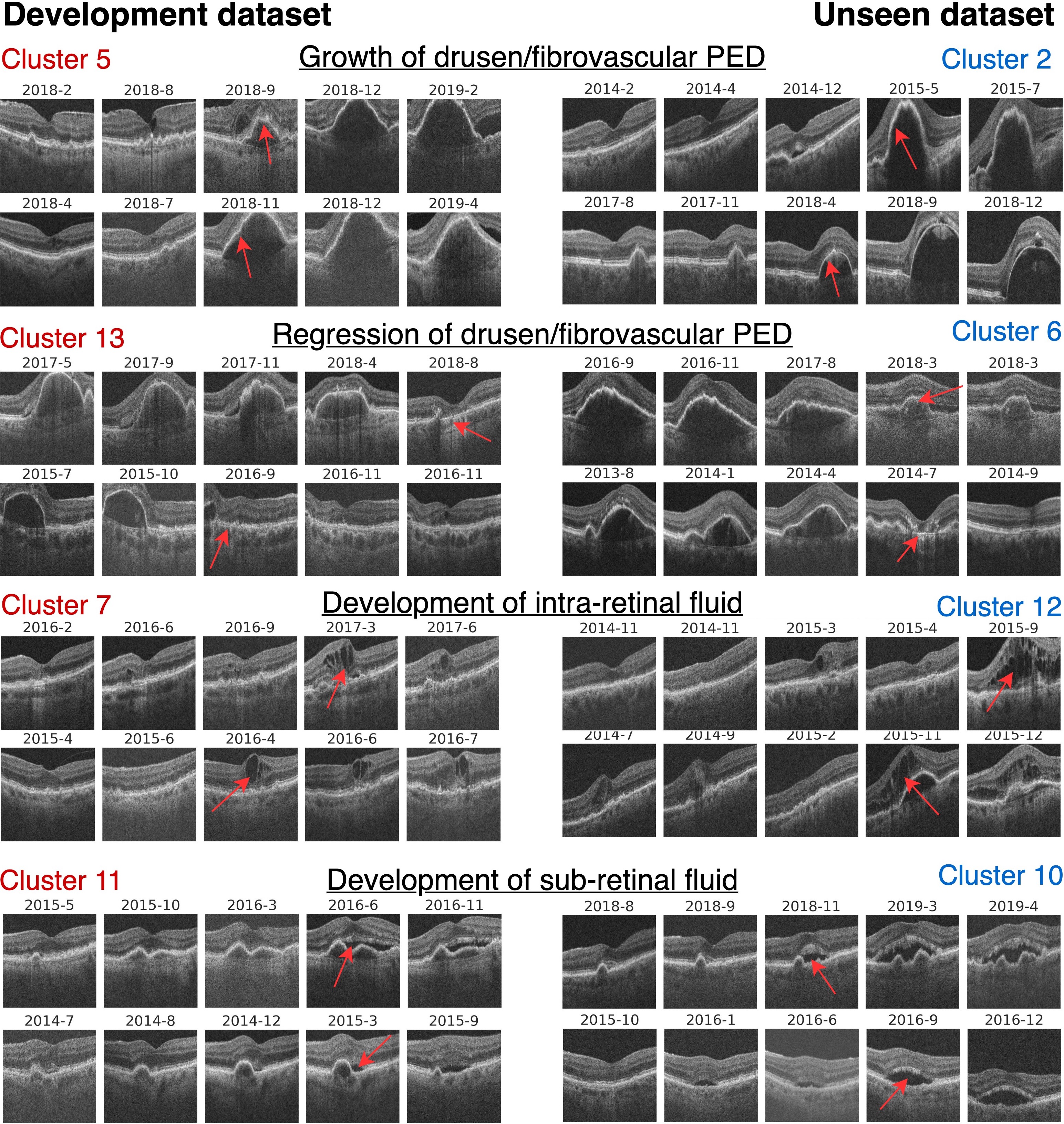}

    \caption{We show four clusters from the \textit{Development dataset} (left half) and the equivalent clusters in the \textit{Unseen dataset} (right half). Ophthalmologists identified clusters capturing the same progression dynamics in both datasets, providing clinical interpretations (underlined). Clusters show two representative sub-trajectories originating from different patients, each containing five longitudinal images with the time and location of greatest progression marked by arrows.
    }
    \label{fig:notable_clusters}
\end{figure}

\subsection{Qualitative and quantitative evaluation of clusters}
Initially, we tune the hyperparameters, $\lambda$, $\phi$ and $K$, on the \textit{Development dataset} by heuristically selecting values that result in higher uniformity between sub-trajectories within each cluster. Expert ophthalmologists then review the clusters, interpreting and summarising any consistently observed temporal dynamics. Next, we apply the method to the \textit{Unseen dataset}, using the same hyperparameters. Ophthalmologists then review these clusters and confirm whether they capture the same temporal biomarkers observed in the \textit{Development dataset}.\\
In addition to the qualitative evaluation, we also validate the utility of our clusters as biomarkers that stratify risk of disease progression. We test this by predicting the time until conversion to late AMD and its subtypes, CNV and cRORA. Additionally, we predict current visual acuity. 
To use our cluster assignments for these tasks, we define a probability distribution over the $K$ clusters and fit a linear regression model. Similarly, we fit an equivalent linear regression model to the static biomarkers from the established grading system detailed in section \ref{method:datasets}. We also include a demographic baseline that combines the patient's age, with which onset of AMD is highly correlated, and sex.
In each case we reserve a random subset of 80\% of the sub-trajectories for the train set and the remaining 20\% for the test set, while ensuring a patient-wise split. We repeat this using 10-fold cross-validation to find the overall performance. Finally, we repeat the entire process, starting from sub-trajectory extraction, followed by clustering and then regression experiments, using 7 random seeds and report the means and standard deviations. Additional information on this pipeline is provided in the supplementary material.

\section{Experiments and results}

\subsection{Sub-trajectory clusters are candidate temporal biomarkers}
By first applying our method to the \textit{Development dataset} we found that using $\lambda=0.75$, $\phi=0.75$ and $K=30$, resulted in the most uniform and homogeneous clusters while still limiting the total number of clusters to a reasonable amount. Achieving the same cluster quality with smaller values of $\phi$ required many more clusters in order to encode all combinations of possible start and end disease states. The expert ophthalmologists remarked that many of the identified clusters capture dynamics that have previously been linked to the progression of AMD, even though they are currently not included in any clinical grading system. Using the same hyperparameters our method generalised to the \textit{Unseen dataset} which yielded clusters with equivalent dynamics and quality (see Figure \ref{fig:notable_clusters}).\\
In both datasets our method differentiated between time series showing growth of pigment epithelial detachments (PEDs) and their eventual regression (resulting in atrophy and conversion to late dry AMD). It also separated development of intraretinal from subretinal fluid, or CNV (conversion to late wet AMD). To see more cluster interpretations we refer to the supplementary material.

\begin{table}[h!]
\centering
\caption{Temporal clusters were comparable to the established clinical grading systems for AMD in predicting risk of future disease, shown by reduced MAE in years.}
\vspace{0.05in}
\begin{tabular}{rcccc}
\cline{2-5}
\multicolumn{1}{c}{}                        & \multicolumn{4}{c}{\textbf{Development dataset}}       \\ \cline{2-5} 
\multicolumn{1}{c}{}                        & \multicolumn{1}{c|}{\begin{tabular}[c]{@{}c@{}}Time to\\ Late AMD $\downarrow$\end{tabular}} & \multicolumn{1}{c|}{\begin{tabular}[c]{@{}c@{}}Time to\\ CNV $\downarrow$\end{tabular}} & \multicolumn{1}{c|}{\begin{tabular}[c]{@{}c@{}}Time to\\ cRORA $\downarrow$\end{tabular}} & \begin{tabular}[c]{@{}c@{}}Current\\ visual acuity $\downarrow$\end{tabular} \\ \hline
\multicolumn{1}{r|}{Demographic}            & \multicolumn{1}{c|}{0.756$\pm$0.010}                      & \multicolumn{1}{c|}{0.822$\pm$0.012}                & \multicolumn{1}{c|}{0.703$\pm$0.028}       & 0.381$\pm$0.007                          \\
\multicolumn{1}{r|}{Current grading system} & \multicolumn{1}{c|}{0.757$\pm$0.010}                      & \multicolumn{1}{c|}{0.819$\pm$0.012}                & \multicolumn{1}{c|}{0.685$\pm$0.035}       & 0.367$\pm$0.008                          \\ \hline
\multicolumn{1}{r|}{Temporal clusters}      & \multicolumn{1}{c|}{\textbf{0.746$\pm$0.011}}            & \multicolumn{1}{c|}{\textbf{0.772$\pm$0.013}}       & \multicolumn{1}{c|}{\textbf{0.619$\pm$0.031}}                                  & \textbf{0.350$\pm$0.009}                  \\
\multicolumn{1}{l}{}                        & \multicolumn{1}{l}{} & \multicolumn{1}{l}{}                                & \multicolumn{1}{l}{}                       &      \\ \cline{2-5} 
\multicolumn{1}{l}{}                        & \multicolumn{4}{c}{\textbf{Unseen dataset}}            \\ \hline
\multicolumn{1}{r|}{Demographic}            & \multicolumn{1}{c|}{1.343$\pm$0.027}                     & \multicolumn{1}{c|}{1.241$\pm$0.017}                & \multicolumn{1}{c|}{\textbf{1.216$\pm$0.062}}                                  & 0.188$\pm$0.007                          \\
\multicolumn{1}{r|}{Current grading system} & \multicolumn{1}{c|}{\textbf{1.308$\pm$0.018}}            & \multicolumn{1}{c|}{1.244$\pm$0.022}                & \multicolumn{1}{c|}{1.286$\pm$0.053}       & \textbf{0.177$\pm$0.008}                 \\ \hline
\multicolumn{1}{r|}{Temporal clusters}      & \multicolumn{1}{c|}{1.322$\pm$0.029}                     & \multicolumn{1}{c|}{\textbf{1.235$\pm$0.027}}       & \multicolumn{1}{c|}{1.257$\pm$0.056}       & 0.188$\pm$0.006                         
\end{tabular}
\label{tab:quant}
\end{table}

\subsection{Newly found biomarkers predict conversion to late AMD}

Next, we validated that our clusters are predictive of progression to late AMD. Our clusters were comparable to, and in some cases outperformed, the current widely adopted grading system in predicting risk of conversion (see Table \ref{tab:quant}). In all tasks the standard biomarkers are only marginally more indicative of risk than the patient's age and sex. This experiment confirms that our clusters are related to disease progression.
\newpage
\section{Discussion and conclusion}

Motivated to improve inadequate grading systems for AMD that do not incorporate temporal dynamics we proposed a method to automatically discover biomarkers that are time-dependent, interpretable, and predictive of conversion to late-stage AMD. We applied our method to two large longitudinal datasets, cataloguing 3,218 total years of disease progression. The found time-dependent clusters were subsequently interpreted by expert ophthalmologists. They found them to capture distinct patterns of disease progression that have been previously linked to AMD, but are not currently included in clinical grading systems. Furthermore, we experimentally demonstrated that the found clusters predict conversion to late-stage AMD on par with the established grading system.\\
Both the strong qualitative and quantitative results were obtained using an unseen testing dataset. This motivates the application of our method to other longitudinal datasets in ophthalmology and beyond. While many clusters identified variants of early AMD, some captured periods after conversion to late AMD. This is due to the over-representation of patients with late disease in our datasets. Moreover, by using 2D OCT images we potentially disregard relevant information contained in the full volumetric scans. \\
In the future, biomarkers identified by our method can be further refined by clinicians. We envision that they may inform the next generation of grading systems for AMD that incorporate the temporal dimension intrinsic to this dynamic disease.

\subsubsection{Acknowledgements}
We thank J. Sutton, A.J. Cree and M. Ruddock for their administrative support of this work. Funding: The PINNACLE Consortium is funded by a Wellcome Trust Collaborative Award, “Deciphering AMD by deep phenotyping and machine learning (PINNACLE)”, ref. 210572/Z/18/Z.
\newpage

\bibliographystyle{splncs04}
\bibliography{samplepaper}

\newpage
\section*{Supplementary material}

\newcommand{\gw}[0]{0.23\textwidth}

\begin{figure}[ht!]
    \centering
    \includegraphics[width=0.86\linewidth]{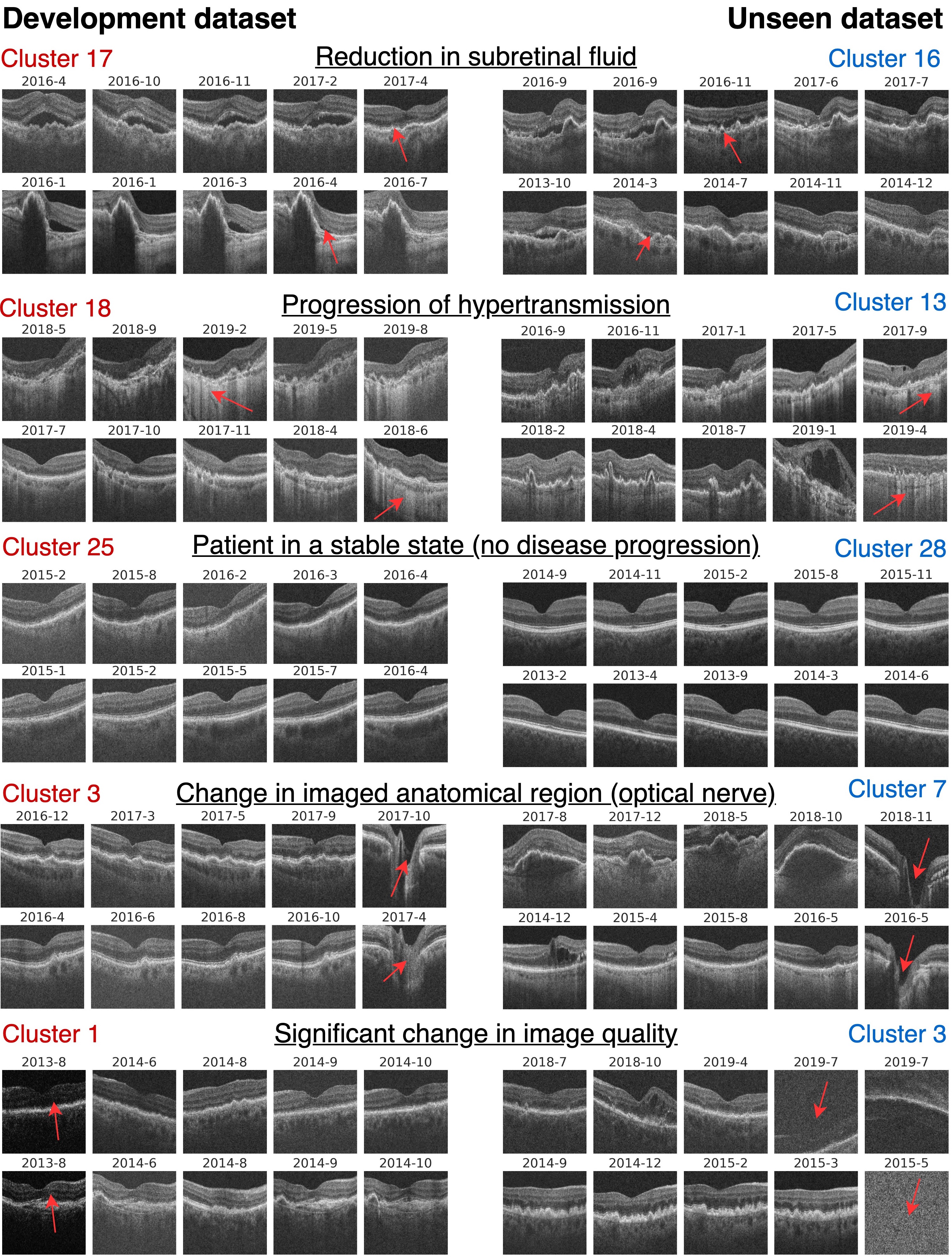}

    \caption{We show five more clusters from the \textit{Development dataset} and \textit{Unseen dataset}. These clusters were found to capture reduction in subretinal fluid (potentially due to treatment), progression of atrophy evidenced by hypertransmission (increased signal under the retina) and a stable disease state with no signs of disease progression. Furthermore, we find two clusters of image artefacts, such as large change in image brightness and quality, and change in anatomical region captured in the image. These clusters were easily identified as artefacts by clinicians and removed from consideration as temporal biomarkers.
    }
    \label{fig:notable_clusters}
\end{figure}

\begin{figure*}[t!]
    \centering
    \includegraphics[width=0.98\linewidth]{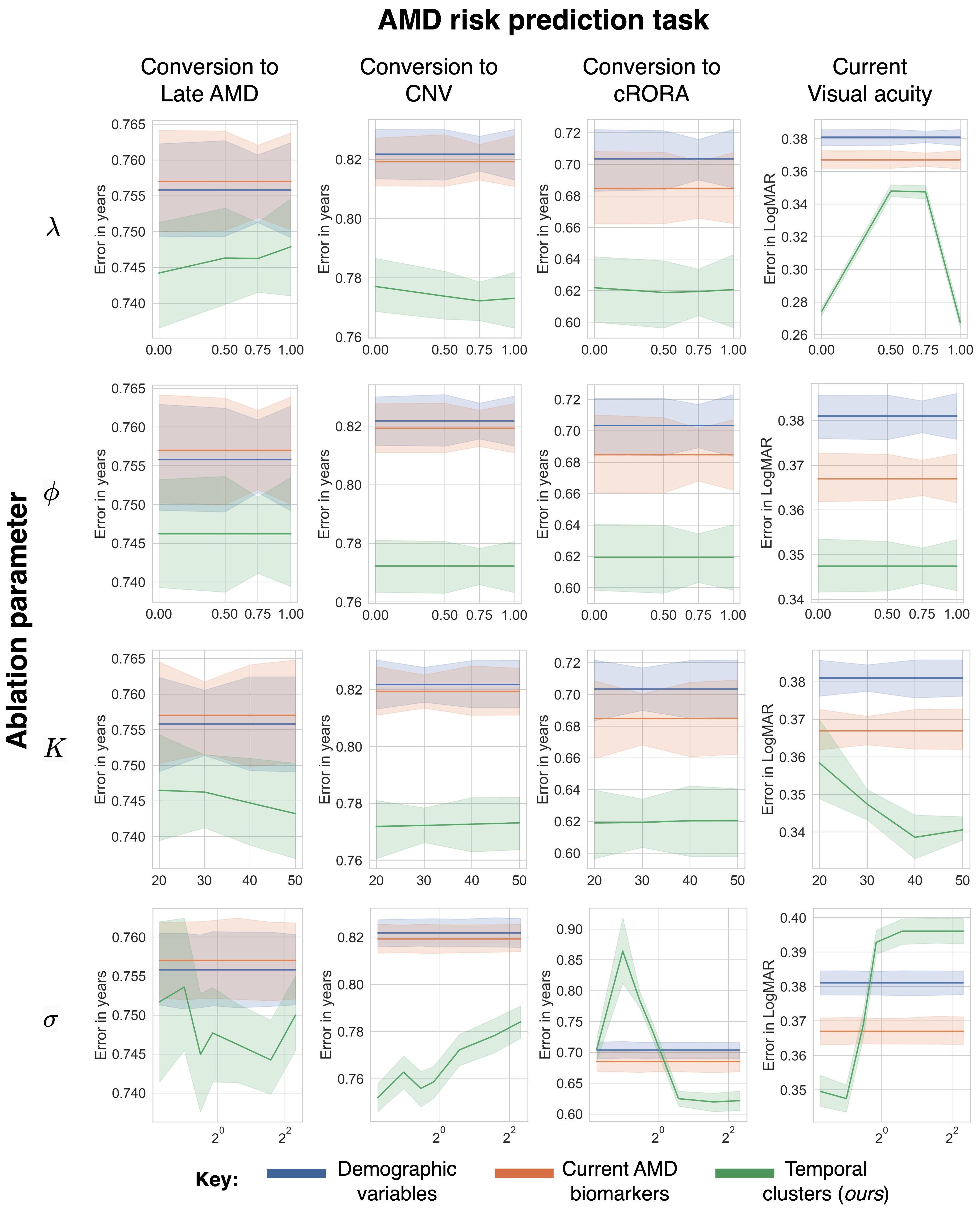}
    
    \caption{Ablation over $\phi$, $\lambda$, $K$ and $\sigma$ parameters (rows) on AMD risk prediction tasks (columns) on the \textit{Development dataset} using seven random seeds. $\sigma$ controls the degree of locality used for risk stratification. For each sub-trajectory we leverage $\mathcal{A}$ to find the average affinity with the members of each cluster. To these $K$ averages we apply a Gaussian kernel, with standard deviation $\sigma$, resulting in $K$ membership probabilities used by the linear risk prediction model. After selecting $\sigma$ for each task on the \textit{Development dataset} we test the same configuration of parameters by applying them to the \textit{Unseen dataset}.}
\end{figure*}

\end{document}